\documentclass[a4paper]{spie}  

\usepackage[]{graphicx}

\usepackage{fleqn,epsfig,color,graphicx,url,psfrag}
\usepackage{amsfonts,amssymb,amsmath,esint}

\newcommand{\myspan}[1]{\mathrm{span}\left\{#1\right\}}
\newcommand{\myset}[1]{\left\{#1\right\}}

\newcommand{\nrb}{N}
\newcommand{\nn}{\mathcal{N}}

\newcommand{\un}{u}
\newcommand{\urb}{\tilde{u}}

\newcommand{\mani}{\mathcal{M}^\nn}

\newcommand{\xn}{X^\nn}

\newcommand{\norm}[2]{\left|\left|#1\right|\right|_{#2}}

\newcommand{\curl}{\mathbf{curl}\;}

\newcommand{\hcurl}{\mathrm{H}\left(\curl,\Omega\right)}

\newcommand{\xitrain}{\Xi_{\mathrm{train}}}

\title{Reduced basis method for source mask optimization}

\author{
Jan Pomplun\supit{\,ab},
Lin Zschiedrich\supit{\,b},
Sven Burger\supit{\,ab},
Frank Schmidt\supit{\,ab},\\
Jacek Tyminski\supit{\,c},
Donis Flagello\supit{\,d},
Nakashima Toshiharu\supit{\,e}
\skiplinehalf
\supit{a}
Zuse Institute Berlin,
Takustra{\ss}e 7,
D\,--\,14\,195 Berlin,
Germany
\smallskip\\
\supit{b}
JCMwave GmbH,
Bolivarallee 22, 
D\,--\,14\,050 Berlin,
Germany
\smallskip\\
\supit{c}
Nikon Precision Inc.,
1399 Shoreway Road,
94002 Belmont,
USA
\smallskip\\
\supit{d}
Nikon Research Corp of America,
12490 N. Rancho Vistoso Blvd.,
85737 Oro Valley, 
USA
\smallskip\\
\supit{e}
Nikon Corporation,
Miizugahara,
Kumagaya,
Japan
}

\authorinfo{
Corresponding author: J. Pomplun\\
URL: http://www.zib.de/nano-optics/\\
Email: pomplun@zib.de
}

\begin{document} 
  \maketitle


\noindent
This paper has been published in Proc.~SPIE Vol. {\bf 7823}
(2010) 78230E, 
({\it Photomask Technology 2010, M. Warren Montgomery; Wilhelm Maurer, Editors})
and is made available as an electronic preprint with permission of SPIE. 
Copyright 2010 Society of Photo-Optical Instrumentation Engineers. 
One print or electronic copy may be made for personal use only. 
Systematic reproduction and distribution, duplication of any material in this paper for a fee or for 
commercial purposes, or modification of the content of the paper are prohibited. \\
{\it Please see original publication for images at higher resolution.}


\begin{abstract}
Image modeling and simulation are critical to extending the limits of leading edge lithography technologies used for IC making. Simultaneous source mask optimization (SMO) has become an important objective in the field of computational lithography. SMO is considered essential to extending immersion lithography beyond the 45nm node. However, SMO is computationally extremely challenging and time-consuming. The key challenges are due to run time vs. accuracy tradeoffs of the imaging models used for the computational lithography. 

We present a new technique to be incorporated in the SMO flow. This new approach is based on the reduced basis method (RBM) applied to the simulation of light transmission through the lithography masks. It provides a rigorous approximation to the exact lithographical problem, based on fully vectorial Maxwell's equations. Using the reduced basis method, the optimization process is divided into an offline and an online steps. In the offline step, a RBM model with variable geometrical parameters is built self-adaptively and using a Finite Element (FEM) based solver. In the online step, the RBM model can be solved very fast for arbitrary illumination and geometrical parameters, such as dimensions of OPC features, line widths, etc. This approach dramatically reduces computational costs of the optimization procedure while providing accuracy superior to the approaches involving simplified mask models. RBM furthermore provides rigorous error estimators, which assure the quality and reliability of the reduced basis solutions. 

We apply the reduced basis method to a 3D SMO example. We quantify performance, computational costs and accuracy of our method.
\end{abstract}


\section{Introduction}
The importance of numerical simulations in the production process of integrated curcuits has grown rapidly over the past years. Accuracy requirements for lithography and scatterometry simulations thereby increase hand in hand with decreasing feature sizes and fabrication tolerances. In many application areas like design and optimization of photomasks or scatterometry, rigorous electromagnetic near-field computations have to be carried out for quantitatively correct results. With mask structures much smaller than the wavelength of incident light, approximative models like Kirchhoff do not give the correct physical solution. Usage of rigorous electromagnetic field (EMF) solvers, e.g., based on Finite Elements (FEM), Finite Difference Time Domain Methods (FDTD), or Rigorous Coupled Wave Analysis (RCWA), however comes at the price of higher computational costs. 

When solving inverse problems for scatterometric parameter reconstruction or numerical determination of an optimal mask layout, not only a single simulation has to be carried out, but multiple EMF solutions have to be computed. Then high computational times for a single forward solutions can make the complete simulation task infeasible. 

However, above applications have a specific feature. Usually the same basic layout has to be simulated multiple times for different values of geometrical parameters, e.g., line width or height, absorber edge angle, etc. The reduced basis method \cite{Pomplun2008a,Pat07a,ROZ08,ZHE08,POM09b} can be applied to this setup. The solution process of the parametrized system is decomposed into an offline step which has to be performed only once and a fast online step. In the offline step the reduced basis is built self-adaptively by solving the underlying model rigorously several times. The full model is then projected onto the reduced basis. In the online step the assembled reduced system can be solved in the order of seconds independent on the size of the original problem. Furthermore, methods from the well established field of a posteriori error estimation of Finite Element theory \cite{AIN00} can be applied to the reduced basis method to assure reliability of the computed output and also for construction of the reduced basis \cite{POM09a,CHE08}.

In the following we start with the description of an exemplary source mask optimization problem. We then give the key ideas and features of the reduced basis method, which is used for efficient solution of the SMO project in the final section. There we also perform a convergence analysis of our method quantifying its performance.

\section{Source mask optimization}
\label{sec:smo}
In this section we describe a source mask optimization problem, which we want to solve with the reduced basis method. As explained in the abstract, source mask optimization is an important objective for realization of ICs of present and future technology nodes. In SMO the shape of the photomask is not optimized for a fixed source like, e.g., in conventional optical proximity correction \cite{POM09c}, but the source itself is variable and included into the optimization process. The goal is to find an optimal source together with the optimal mask simultaneously. The combined degrees of freedom of source and mask enlarge the dimension of the parameter space and therewith also the computational costs for determination of the optimal parameter values significantly. Furthermore, since structures on the mask are getting much smaller than the wavelength of incident light, approximative models like Kirchhoff are not valid and rigorous solutions have to be computed during optimization. SMO is therefore considered computationally extremely challenging and time consuming.

\begin{figure}
(a)\hspace{9cm}(b)\vspace{1.0cm}\\
\includegraphics[height=7cm]{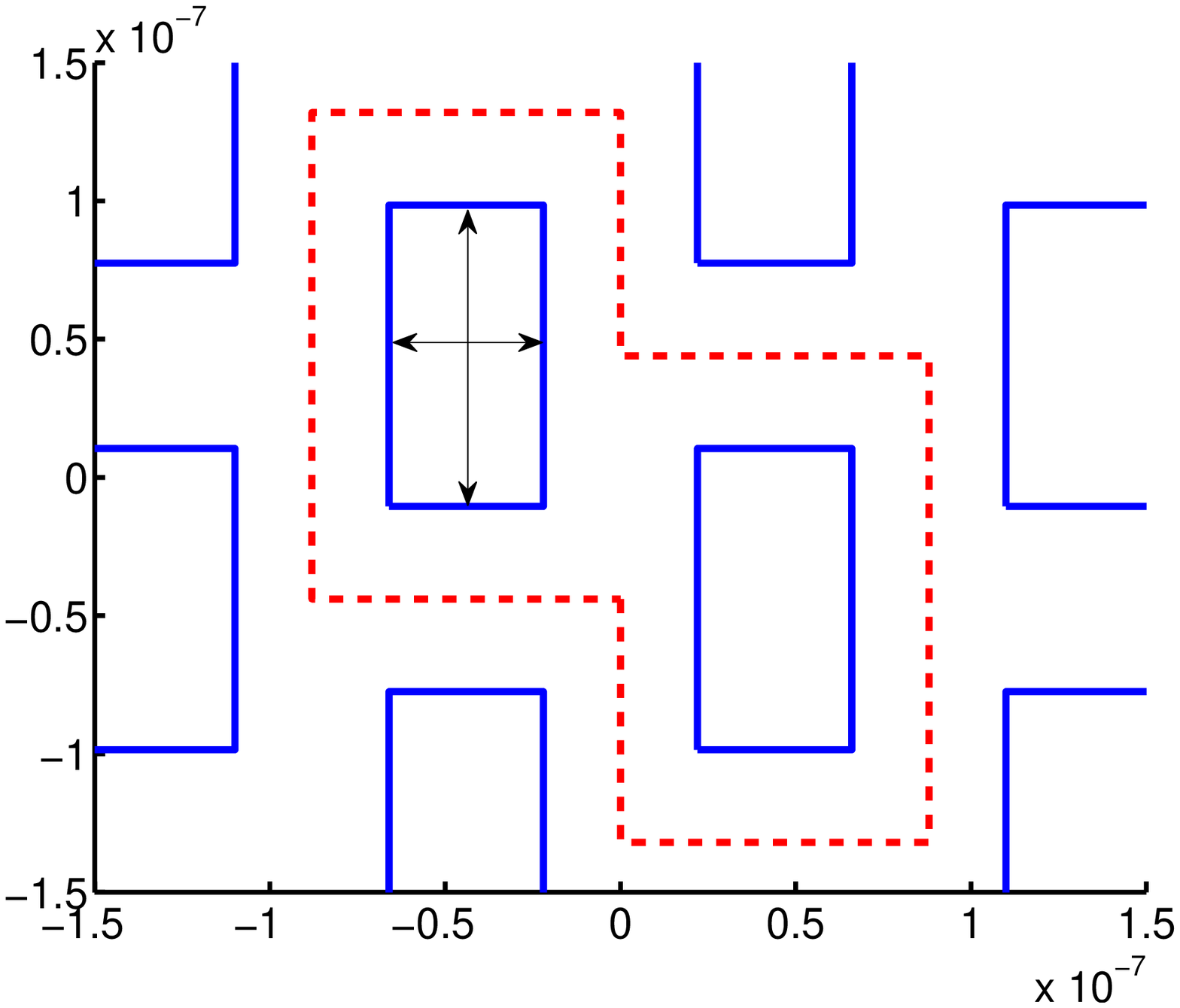}\hfill
\includegraphics[height=7cm]{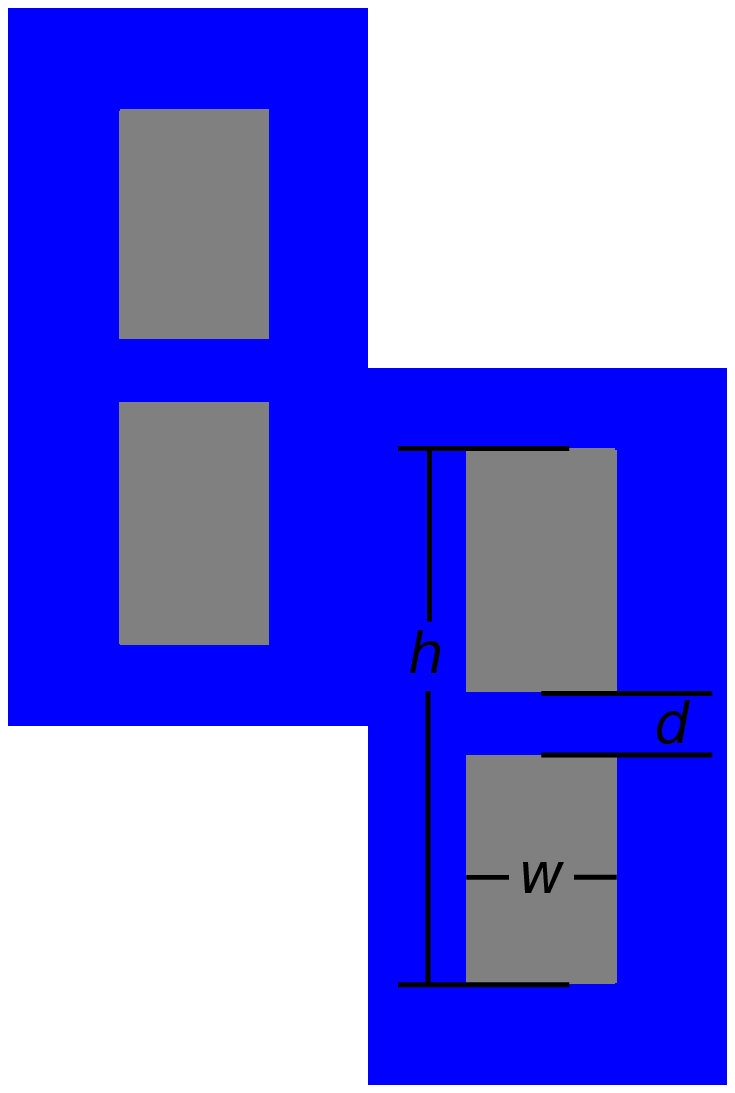}\\
(c)\hspace{9cm}(d)\vspace{1.0cm}\\
\includegraphics[height=6cm]{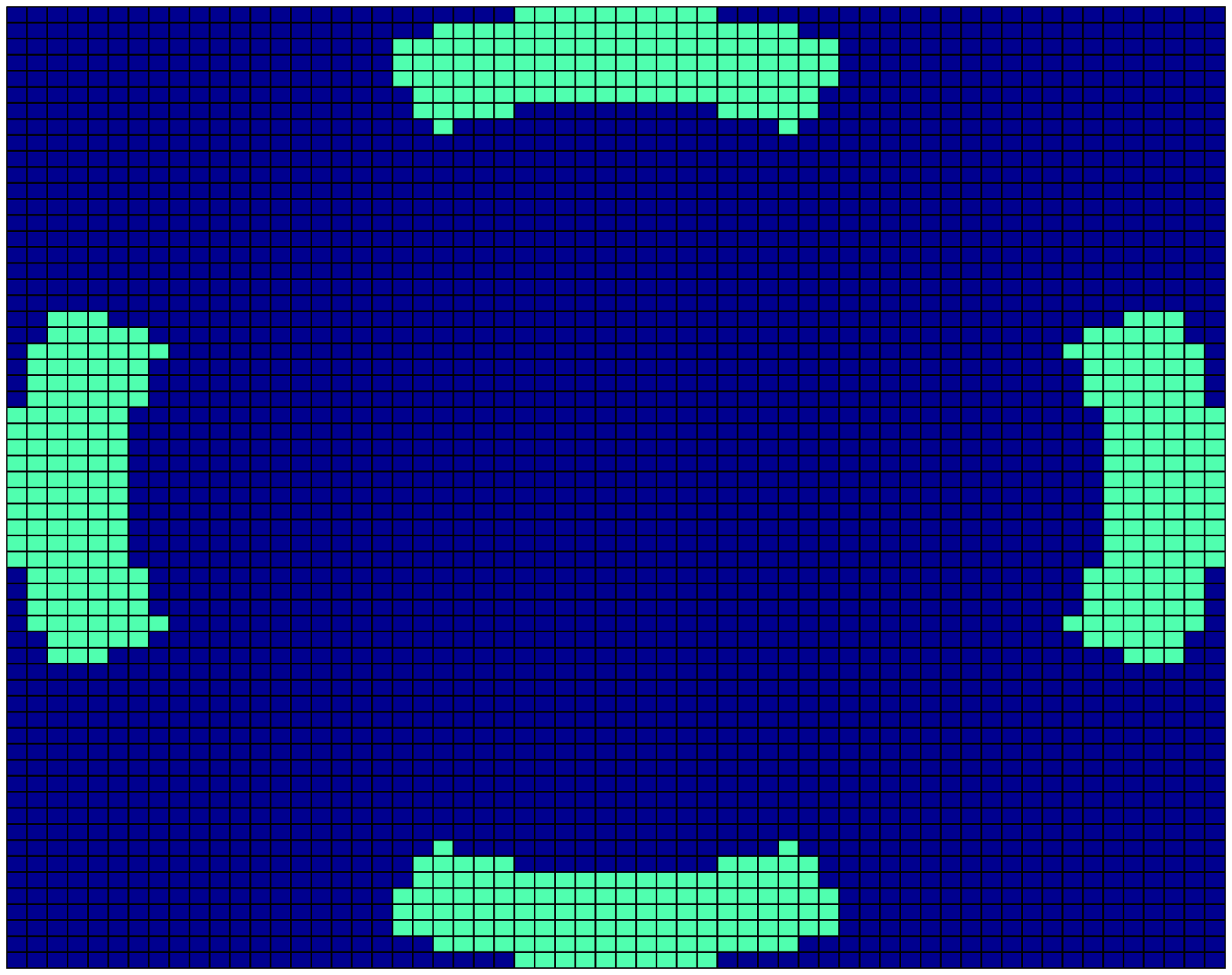}\hfill
\includegraphics[height=6cm]{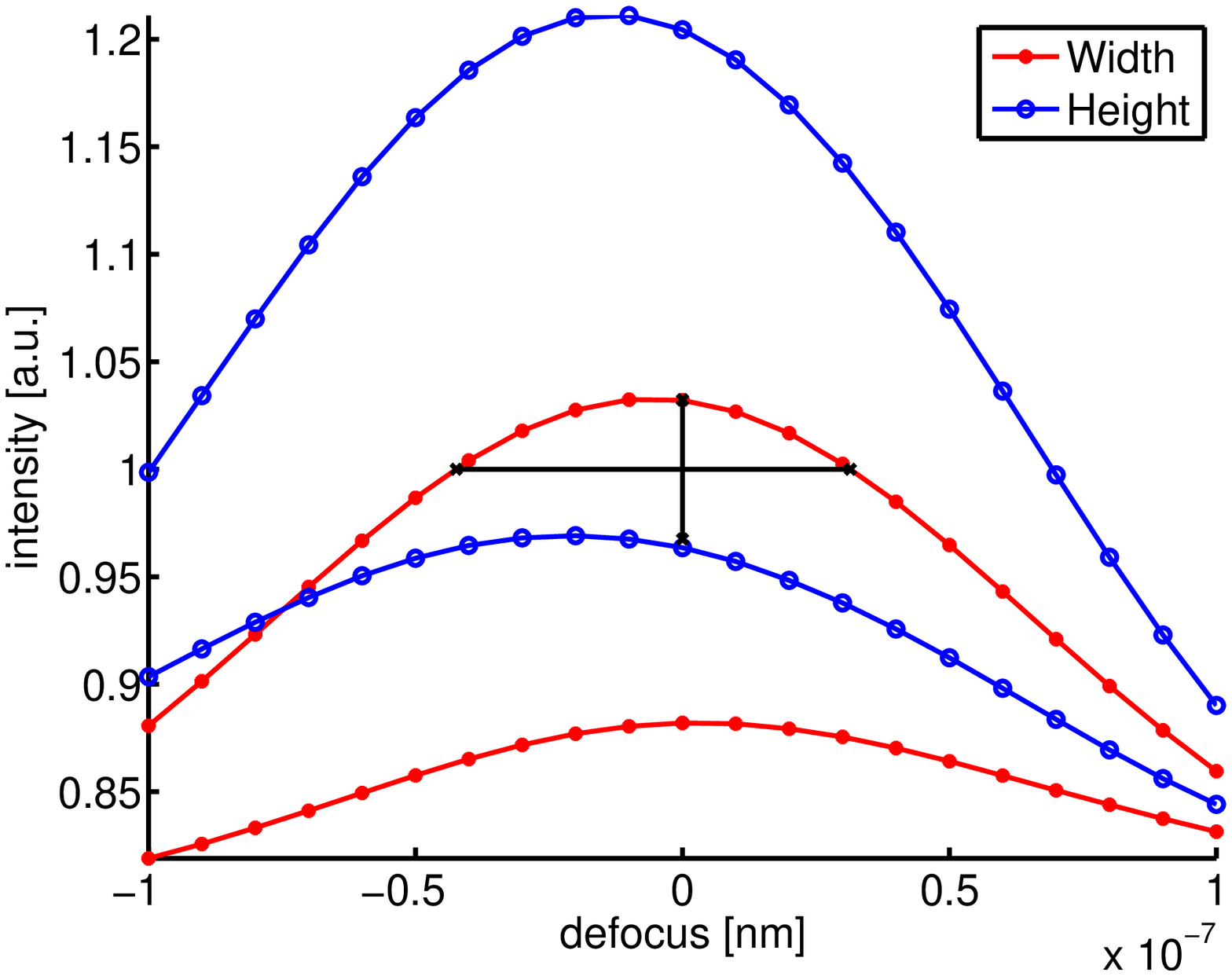}
\caption{\label{fig:target}}
(a) Target cutting hole structure on wafer; dashed line denotes unit cell of layout; arrows denote position for determination of hole width and height for process window definition. (b) Top image of a unit cell of the periodic photomask. (c) Pixelated quadrupole illumination. (d) Sample combined process window for target width and height of cutting hole ($\pm 10\%$).  Vertical black line: height of the process window at zero defocus; horizontal black line: width of the process window at midpoint of vertical line. For optimization, the size of the process window was taken as the product of the length of the sketched black lines.
\end{figure}

The optimization problem we want to tackle is an exemplary cutting lithography step \cite{NAK09}, used for hole patterning of a line-and-space pattern, see Fig.~\ref{fig:target}(a). Our goal is maximization of the process window. Parameters of the system are given in Table \ref{tab:paras}.
\begin{table}
\centering
\begin{tabular}{cc}
parameter & value\\
\hline
wavelength $\lambda$& 193nm\\
NA & 1.35\\
magnification & 4x\\
$n_{\mbox{substrate}}$ & 1.564\\
$k_{\mbox{substrate}}$ & 0\\
$n_{\mbox{absorber}}$ & 0.84 \\
$k_{\mbox{absorber}}$ & 1.65
\end{tabular}
\caption{\label{tab:paras}Parameters of source mask optimization example.
}
\end{table}

SMO for this setup, based on Kirchhoff's method has been presented in \cite{NAK09}. 

\begin{figure}
(a)\hspace{4cm}(b)\hspace{4cm}(c)\hspace{4cm}(d)\hfill\\
\includegraphics[height=5.7cm,clip]{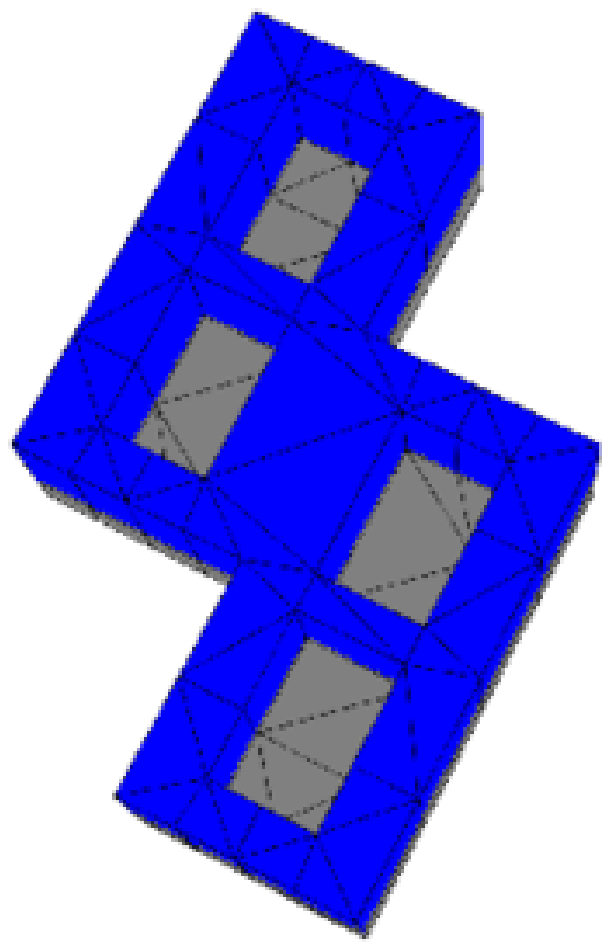}\hfill
\includegraphics[height=5.7cm,clip]{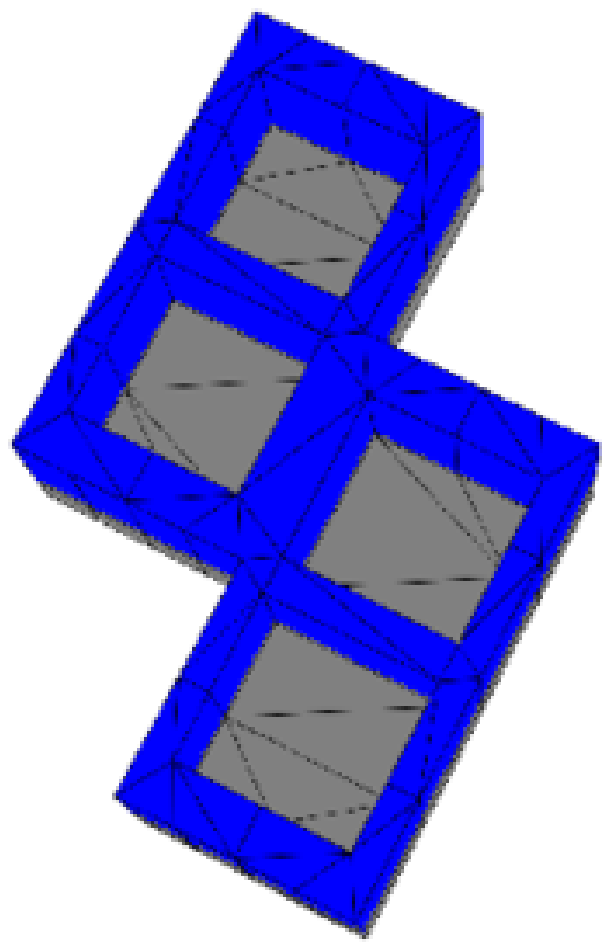}\hfill
\includegraphics[height=5.7cm,clip]{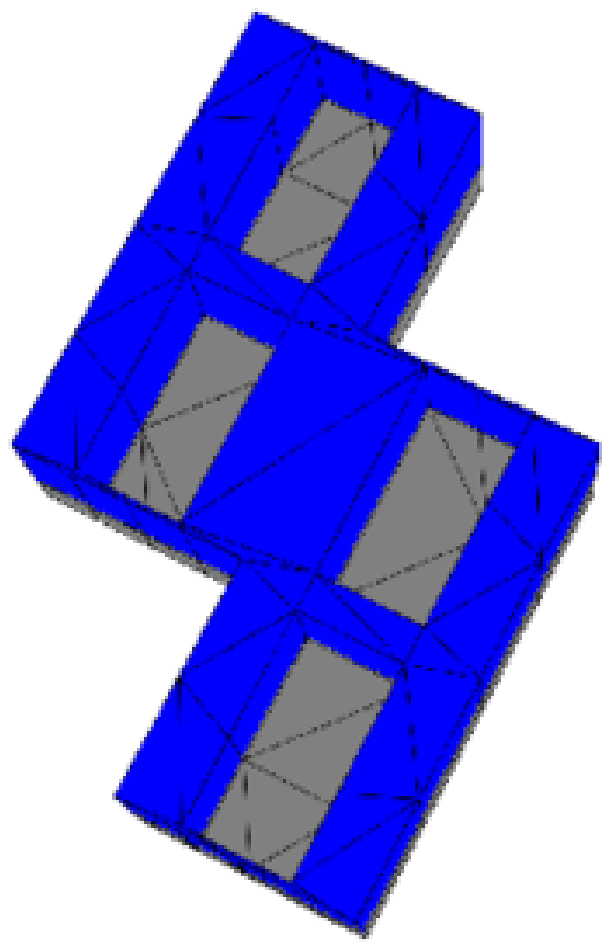}\hfill
\includegraphics[height=5.7cm,clip]{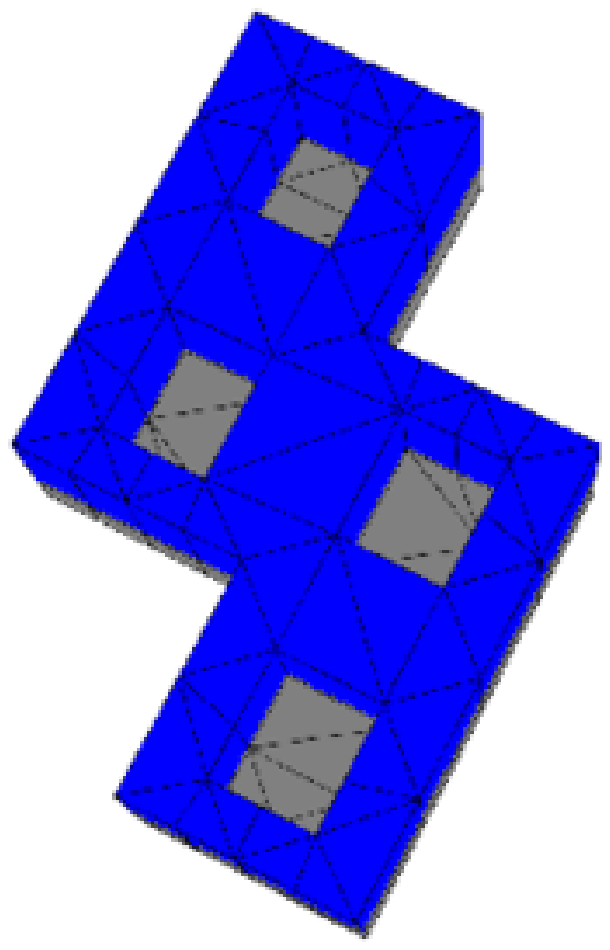}
\caption{\label{fig:geoPara}
Geometrically parametrized mask. (a) Basic layout. (b) Variable hole width $w$. (c) Variable hole height $h$. (d) Variable assist feature $d$. (c.f. Fig.~\ref{fig:target}(b))}
\end{figure}

\begin{figure}[ht]
\vspace{5mm}
(a)\hspace{9cm}(b)\vspace{1.0cm}\\
\includegraphics[height=6cm,clip]{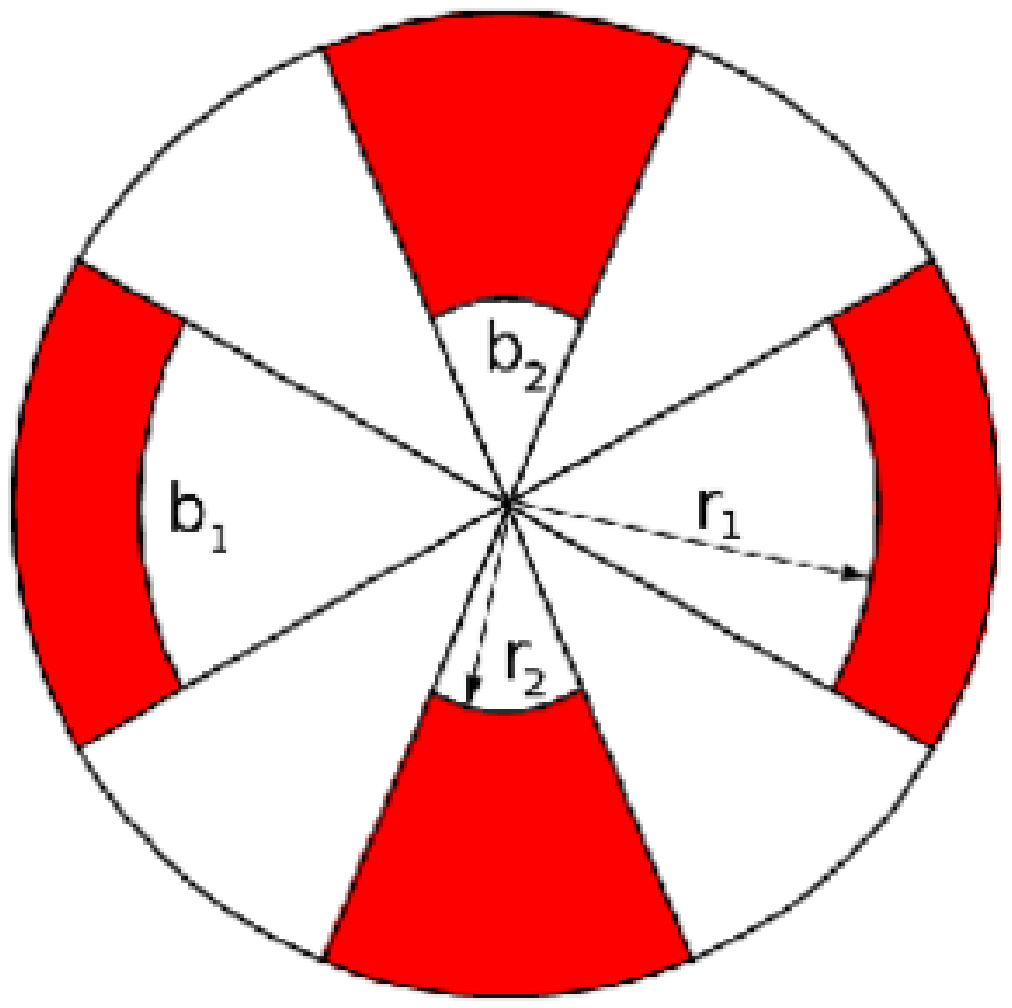}\hfill
\includegraphics[height=6cm,clip]{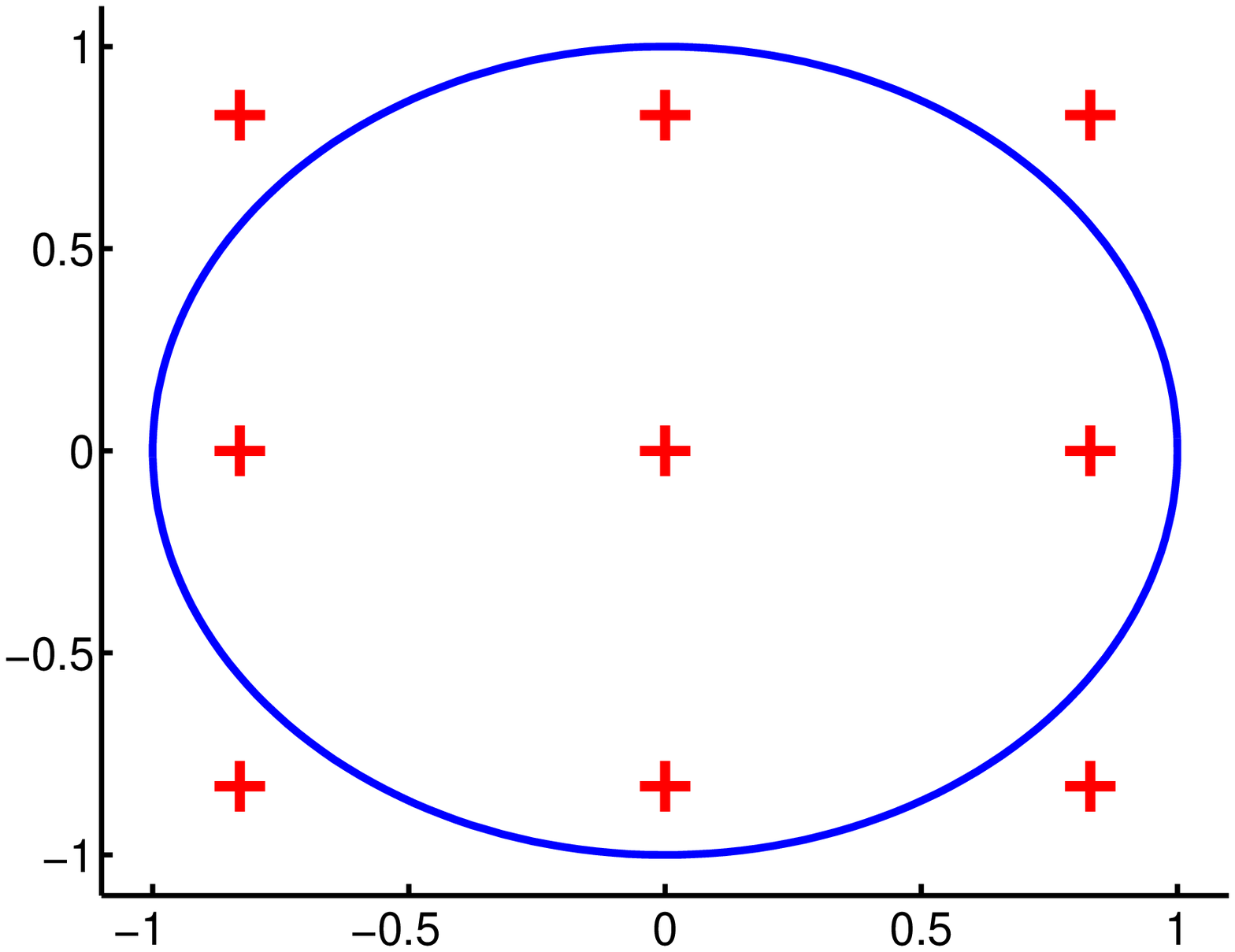}
\caption{\label{fig:geoSource}(a) Quadrupole illumination with four variable parameters; $b_1,b_2$: segment opening angles, $r_1,r_2$ inner segment radii. (b) Incoming directions in the $\sigma$-plane. Crosses denote incoming directions with equal Bloch-periodicity which are computed simultaneously in the Finite Element and reduced basis simulation.
}
\end{figure}

The photomask and parametric illumination of this example are shown in Fig.~\ref{fig:target}(b),(c). For determination of the process window, given exemplary in Fig.~\ref{fig:target}(d), we measure width and height of the holes created on the wafer. Now we parametrize the mask and illumination. The mask topology is described by three geometrical parameters visualized in Fig.~\ref{fig:geoPara}. 
As variable source we introduce four parameters to the quadrupole illumination, depicted in Fig.~\ref{fig:geoSource}. Freeform optimization of the source leads to no principal changes of the reduced basis flow, however in this demonstrative study we only consider a parametric source. Each point in the pixelated representation of the illumination stands for two plane waves with orthogonal polarization to model unpolarized light.

Our optimization task is the maximization of the process window in the seven-dimensional parameter space. For determination of a single process window, we first compute the electromagnetic near-field for fixed source and mask geometry. From this we deduce far field coefficients, aerial image, and contour plots on the wafer. From these we compute the process window. Here, computation of the near-field is by far the most expensive step. Discretization with Finite Elements gives a system with $\nn=341440$ unknowns, which results in single CPU computational times of the order of $10,000\,$sec for 18 plane wave sources, consisting of nine incoming directions times two orthogonal polarizations, see Fig.~\ref{fig:geoSource}(b). All these incoming directions have same Bloch-periodicity, i.e., phase jump from left to right and top to bottom boundary of the computational domain, and can be computed simultaneously in a Finite Element simulation. These sources are used in a local/hybrid \cite{ADA08} Hopkins approach to obtain the far field coefficients for the continuous (numerically pixelated) off-axis illumination. Thereby, the far field of an arbitrary incoming field is determined from the Finite Element solution which has the closest incoming direction.

During optimization, several hundred of these forward solutions have to be computed, which makes the task basically infeasible, due to long run-times. 

In the following we describe the reduced basis method which will dramatically decrease the computational time for the optimization process. It replaces the near-field solution by a quasi-rigorous solution of controlled quality, which is obtained from a reduced model. We will give the key features and ideas of the reduced basis method in the following section. A detailed mathematical discussion can be found in \cite{POM09a}.

\section{Reduced basis method}
For rigorous numerical solution of Maxwell's equations, a discretization scheme like FEM, FDTD, or RCWA has to be applied. This results in a system of algebraic equations for the degrees of freedom which represent the electromagnetic field. Especially in 3D, the computational times for determining the solution to the discretized equations can get very large. For inverse or optimization problems a large number of such solutions have to be computed for different parameter values of a parametrized system. This makes the solution of such real-time or many-query problems extremely time-consuming, inconvinient for a user, and often infeasible.

The idea of the reduced basis method is the usage of precomputed solutions in an actual application. In a relatively expensive offline step a reduced model is constructed by multiple solution of the original problem for different parameter values. The assembled reduced model can then be solved very fast in the online phase. During an actual application, only the online step is evoked. At first glance, above strategy seems comparable to a table based ansatz where solutions are computed for different parameter values offline followed by interpolation online. However, in the following we will see that the reduced basis method has great advantages regarding offline computational times and reliability of the solution.

For application of the reduced basis method, a discretization with Finite Elements is the method of choice and builds its foundation. With Finite Elements, Maxwell's equations are transformed into a sparse matrix equation:
\begin{align*}
A\un=b,
\end{align*}
where $A$ is the sparse system matrix, $u$ the solution vector, and $b$ the right hand side, which includes source fields. we call this system our ``truth approximation''.
For real-world problems the dimension $\nn$ of this system can be up to several millions, which leads to large computation times. Now consider a geometrically parametrized problem as depicted in Fig.~\ref{fig:geoPara}. Since the domain of interest is parameter-dependent, the Finite Element system and the solution $u$ will also become parameter- ($\nu$-)dependent:
\begin{align}
\label{eq:max1}
A_{\nu} \un_{\nu}=b_{\nu}.
\end{align}
During optimization we want to solve this equation for a large number of values for $\nu$, which are elements of a parameter space $D$.

\begin{figure}[ht]
\centering
\psfrag{mani}{$\mani$}
\psfrag{nn}{$\xn$}
\psfrag{s1}{$\urb(\nu_{1})$}
\psfrag{s2}{$\urb(\nu_{2})$}
\psfrag{s3}{$\urb(\nu_{3})$}
\includegraphics[width=7cm]{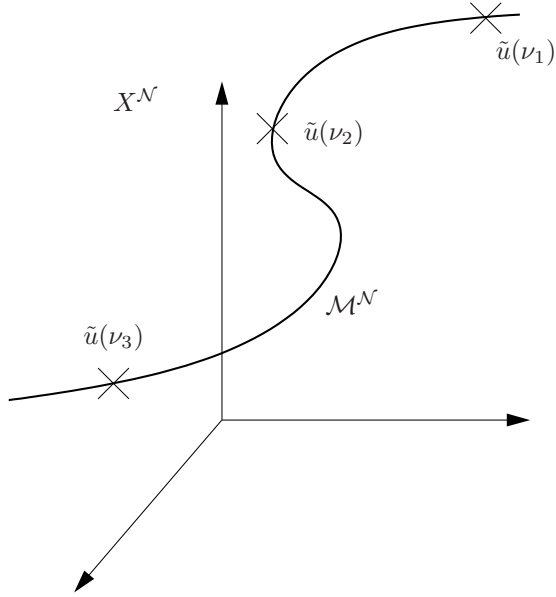}\hfill
\caption{\label{fig:solMani}Sub-manifold of possible solutions $\mani$ in Finite Element space $\xn$ for a single parameter $\nu$. Three snapshot solutions $\urb(\nu_{i})$ are depicted.}
\end{figure}
The basic idea of the reduced basis method is depicted in Fig.~\ref{fig:solMani}. Suppose we have only one parameter which is varied. Then the solution $\un_{\nu}$ will stay on a (maybe complicated) one-dimensional sub-manifold $\mani$ in the high-dimensional Finite Element space. Suppose we would know a subspace $U$ ($\nn\times\nrb$-matrix) of the Finite Element space, which gives a good approximation to the sub-manifold of all possible solutions:
\begin{align*}
U\approx\mani.
\end{align*}
Then we can represent the solution $\un_{\nu}$ of \eqref{eq:max1} by a linear combination of elements of $U$:
\begin{align}
\label{eq:ansatz1}
\un_{\nu}\approx U\lambda_{\nu}=\urb_{\nu},
\end{align}
where $\lambda_{\nu}$ is a $\nrb$ dimensional coefficient vector. Now we insert ansatz \eqref{eq:ansatz1} into the discretized Maxwell's equations \eqref{eq:max1} and project the resulting system onto $U$. This gives:
\begin{align}
\label{eq:maxRB1}
\left[U^H A_{\nu} U\right]\lambda_{\nu}=U^H b_{\nu},
\end{align}
where the superscript $H$ denotes the Hermitian transpose. This system of equations only has $\nrb$ unknowns, the coefficient $\lambda_{\nu}$, and can be solved very fast if $\nrb$ stays moderate $O(100)$. Equation~\eqref{eq:maxRB1} is the reduced basis system, $U$ the reduced basis, and $\urb_\nu$ the reduced basis solution. An important issue is the construction of the reduced basis $U$, which we explain in the following section. Before that, we will comment on the features of the reduced basis method and compare its properties to table based approaches.

In the online step of a table based ansatz, the physics of the underlying system is completely neglected. The output of interest for a specific parameter is simply computed by interpolation from output values of other parameters. In the reduced basis setting Eq.~\eqref{eq:maxRB1} can be interpreted as a scattering problem stated on the space of global ansatz functions $U$ instead of the Finite Element space \cite{POM09a}. This means in the online step we still solve Maxwell's equations rigorously, but only on a reduced solution space. As we will see, this leads to a very fast exponential convergence of the method with increasing dimension of $U$ and is a way to overcome the ``curse of dimensionality'' for high-dimensional parameter spaces which is a serious problem for all table based approaches.

Another great advantage of the reduced basis method is the possibility to estimate the error of a reduced basis solution, without computing the exact Finite Element solution. This is the field of a posteriori error estimation, well investigated in the field of Finite Elements. The basic idea is an evaluation of the residuum of a reduced basis solution, which is defined by:
\begin{align}
\label{eq:res1}
r_{\nu}=b-A_{\nu} \urb_{\nu}.
\end{align}
From discretized Maxwell's equations \eqref{eq:max1} we see that the residuum vanishes for the exact solution $\un_\nu$. If we insert a reduced basis solution into \eqref{eq:res1} and compute its residuum we can utilize $\norm{r_{\nu}}{}$ as a measure for the error of the solution. We will demonstrate in our numerical example the good performance of this estimator. We can use this estimator to scan the parameter domain and identify regions, where the reduced system still needs improvement. Comparable methods do not exist for table based approaches.

The error estimator is also used for self adaptive construction of the reduced basis approximation. In an application the user only specifies the parameter space $D$ and the error estimator decides how to construct $U$ as will be explained in the following.

\subsection{Construction of the reduced basis space}
The reduced basis space $U$ has to be constructed such that it gives a good approximation to the space of all possible solutions $\mani$ of the parameter domain. In order to explain the construction algorithm, we first define a sequence of hierarchical subsets of the parameter domain $D$. Let $\nu_{i}\in D\,,\;i=1,\dots,\nrb_{\mathrm{max}}$. Then we define:
\begin{align}
  \label{eq:defNestPar}
S_{i}=&\myset{\nu_{1},\dots,\nu_{i}}\,,\quad i=1,\dots,\nrb_{\mathrm{max}}.
\end{align}
These sets have the following property:
\begin{align}
  \label{eq:nestParHier}
S_{1}\subset S_{2}\subset\dots\subset S_{\nrb}\subset\dots\subset S_{\nrb_{\mathrm{max}}}.
\end{align}
The reduced basis space $U_\nrb$ of dimension $\nrb$ is then defined by:
\begin{align}
  \label{eq:defLagSpace}
U_{\nrb}=&\,\myspan{\un(\nu) \mbox{ is a solution to \eqref{eq:max1}}\;|\;\nu\in S_{\nrb}},
\end{align}
hence it is spanned by solutions to the truth approximation for fixed parameter values. These solutions are called snapshot solutions, see Fig.~\ref{fig:solMani}. It is also possible to include first and higher derivatives of the field $\un$ with respect to parameters into the reduced basis space, which leads to so called Taylor and Hermite spaces. Here we notice the expensive ``offline'' costs of the reduced basis method. For construction of a reduced basis space of dimension $\nrb$ the truth approximation, i.e., the full problem, has to be solved $\nrb$ times. However, the reduced basis space only has to be assembled once and furthermore a parallelization of this process is possible.

The final question is how to choose the snapshot parameters $\nu_{i}$ in \eqref{eq:defNestPar}. The following {\it greedy algorithm} \cite{ROZ08} results in reduced basis spaces of small dimension and very good approximation quality.

First we define a training set $\xitrain\in D$ of possible snapshot candidates. From this space we want to chose a number $\nrb$ of snapshot parameters for construction of the reduced basis. The first parameter $\nu_{1}$ is chosen randomly. Then we construct a one dimensional reduced basis approximation corresponding to the snapshot $\urb(\nu_{1})$. Now we evaluate the error estimator for this one dimensional reduced basis approximation on all candidate snapshots in the training sample $\xitrain$ and include the parameter value with the maximum error into the reduced basis because this is supposed to add a maximum of ''new information'' into the reduced basis. Then we have a two dimensional reduced basis and the process is continued iteratively. The process can be stopped, e.g., if a certain maximum dimension is reached or if the error estimator gives sufficiently small bounds over the training set.

\begin{figure}
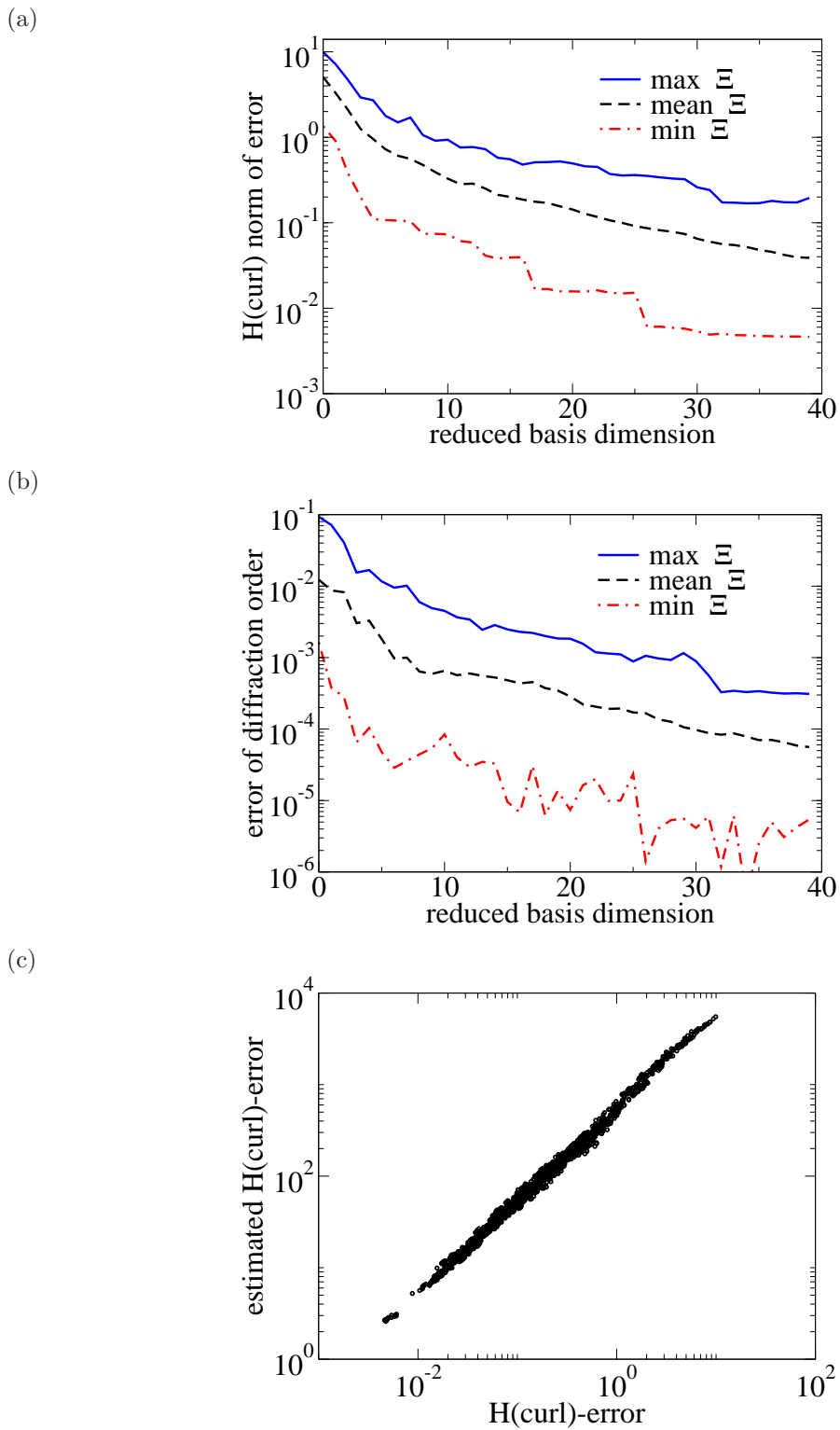

(a)\\
\phantom{aaaaaaaaaaaaaaaaaaa}\includegraphics[width=8.5cm,clip]{fig/hCurl_Dim.eps}\vspace{2mm}\\
(b)\\
\phantom{aaaaaaaaaaaaaaaaaaa}\includegraphics[width=8.5cm,clip]{fig/order-2_Dim.eps}\vspace{2mm}\\
(c)\\
\phantom{aaaaaaaaaaaaaaaaaaa}\includegraphics[width=8.5cm,clip]{fig/hCurl_est.eps}
\caption{\label{fig:convRB} Convergence of reduced basis solutions over a random parameter ensemble $\Xi$; minimum, mean, and maximum errors are shown. (a) Error in $H(\mbox{curl})$-norm of near-field solution, (b) error of diffraction mode. (c) Scatter plot of estimated errors in dependence on true $H(\mbox{curl})$-error over the random parameter ensemble and reduced basis dimensions 1 to 40.
}
\end{figure}
\section{Numerical results}
In this section we perform the source mask optimization described in Sec. \ref{sec:smo} with a reduced basis approximation for the electromagnetic near-field computation. First we specify the geometrical parameters of the mask, whose unit cell is shown in fig. \ref{fig:target}(b). The pitch values are given by $p_x=p_y=704\,$nm. The geometrical parameter space $D$ of the mask was chosen as:
\begin{eqnarray}
\begin{split}
w&\in\left[120;200\right]\,\mbox{nm},\\
h&\in\left[472;632\right]\,\mbox{nm},\\
d&\in\left[24;120\right]\,\mbox{nm},
\end{split}
\label{eq:paraDomain}
\end{eqnarray}
c.f. Fig.~\ref{fig:geoPara}. 

For above parameter space, a reduced basis approximation of dimension 40 was constructed with the described greedy algorithm. Before we look at the optimization results we want to quantify the accuracy of the reduced basis approximation. Therefore, we randomly choose 50 parameters in the parameter domain \eqref{eq:paraDomain} and compute the corresponding exact Finite Element solutions. This defines a random ensemble $\Xi$. Then we compare each of these exact solutions with the reduced basis results for the same parameters and increasing dimension of the reduced basis approximation: in Fig.~\ref{fig:convRB}(a) the errors of the reduced basis near-field approximations in $\hcurl$-norm are shown, Fig.~\ref{fig:convRB}(b) gives the error of a reduced basis diffraction order, which is used for computation of the aerial image. The mean, minimum and maximum errors over the random ensemble $\Xi$ are shown. We observe exponential convergence of the errors with increasing reduced basis dimension. This means the more snapshots we include into the reduced basis system, the more accurate results we obtain. Reduced basis approximation converges to the exact Finite Element solution. The mean error for the output of interest, i.e., the diffraction orders is below $10^{-4}$ for a dimension of only 40 which guarantees reliable results. 

As explained before, the high convergence rate is a major advantage of the reduced basis method. Only few snapshot solutions in the offline phase are necessary to obtain results with very small errors in the online phase. Next we look at the performance of the error estimator. Figure \ref{fig:convRB}(c) shows a scatter plot of the estimated error in dependence on the true error over the random parameter ensemble. Thereby, the error of the near-field solutions given in \ref{fig:convRB}(a) is given in dependence on the estimated error for each of the 50 random parameters and reduced basis dimensions 1 to 40. We observe excellent correlation of both quantities over several orders of magnitude of the true error. This assures reliability of the reduced basis approximation, since the error estimator can be utilized to check if the reduced basis solution has sufficient quality in the whole parameter space. Solution of the reduced basis system of dimension 40 takes approximately 1s. Compared to a full Finite Element simulation, this results in a speed-up factor of 10,000. 

Due to the short online solution time of the reduced basis model, it is now very convenient to perform parameter scans and source mask optimizations. 
\subsection{Source mask optimization}

\begin{figure}
(a)\hspace{8.8cm}(b)\vspace{0.0cm}\\
\includegraphics[width=8.3cm,clip]{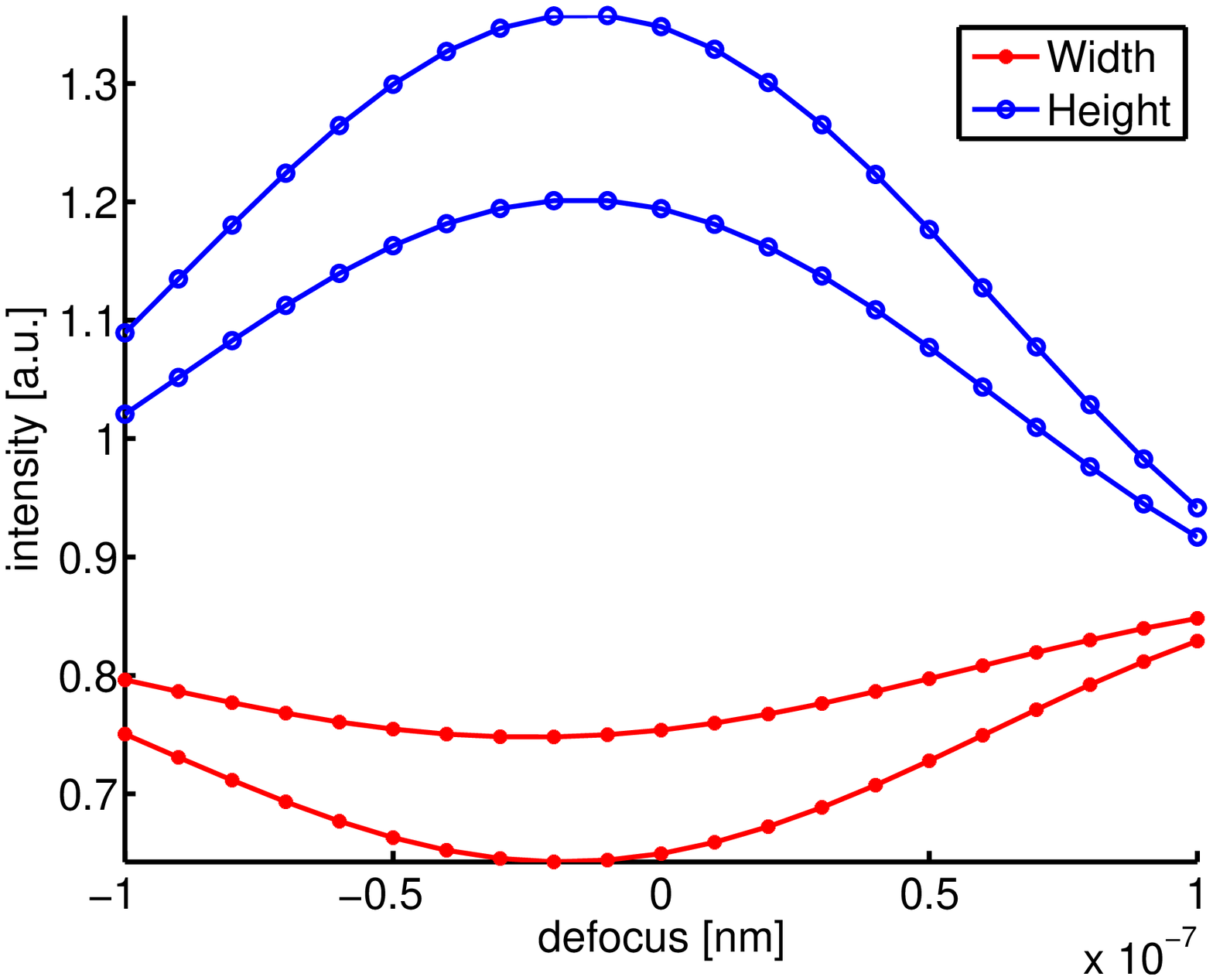}\hfill
\includegraphics[width=8.3cm,clip]{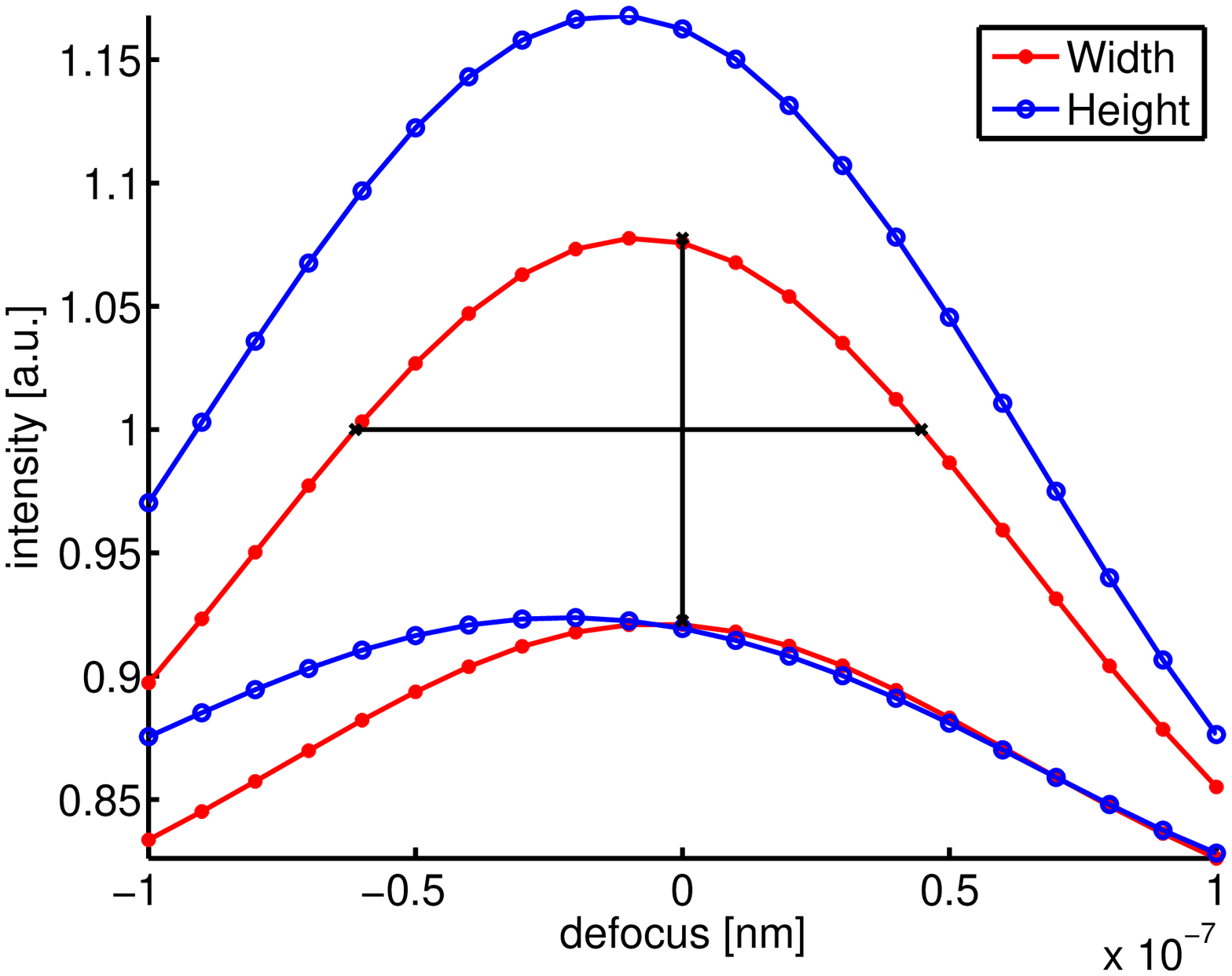}\\
(c)\hspace{8.8cm}(d)\vspace{0.0cm}\\
 \includegraphics[width=7.8cm,clip]{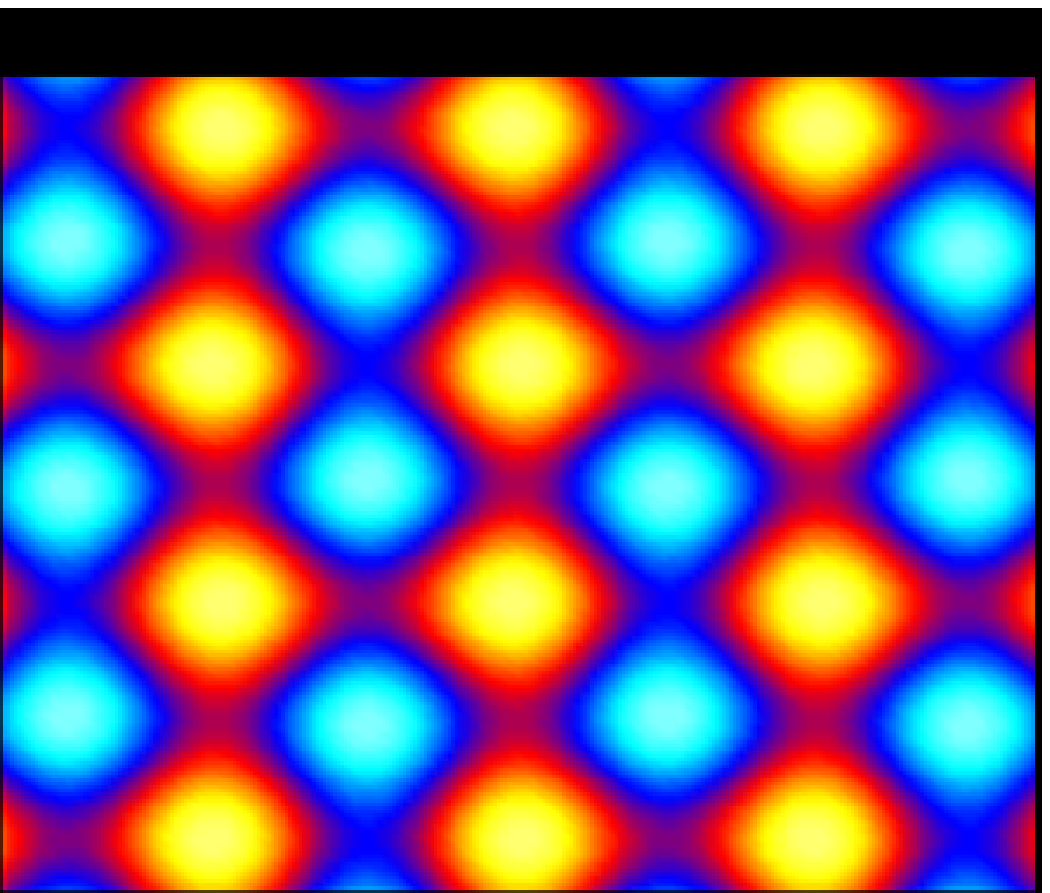}\hfill
 \includegraphics[width=7.8cm,clip]{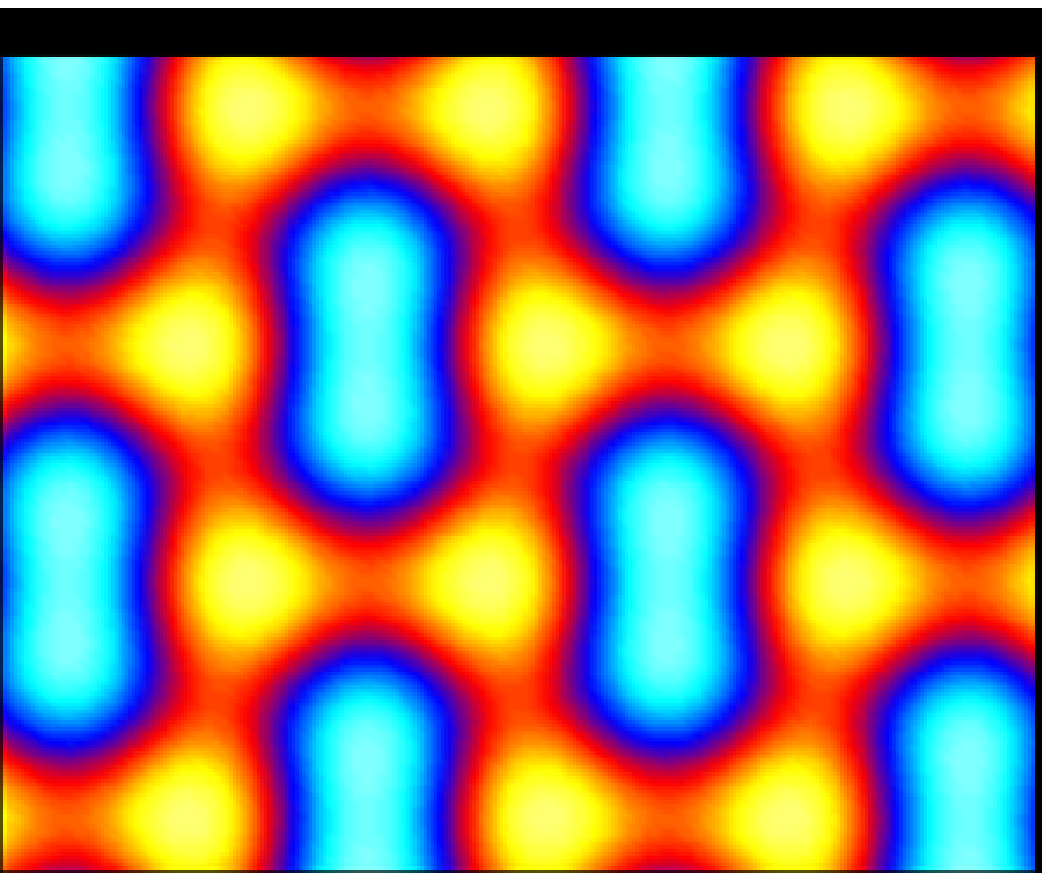}\vspace{2mm}\\
(e)\hspace{8.8cm}(f)\vspace{0.0cm}\\
\includegraphics[width=7.8cm,clip]{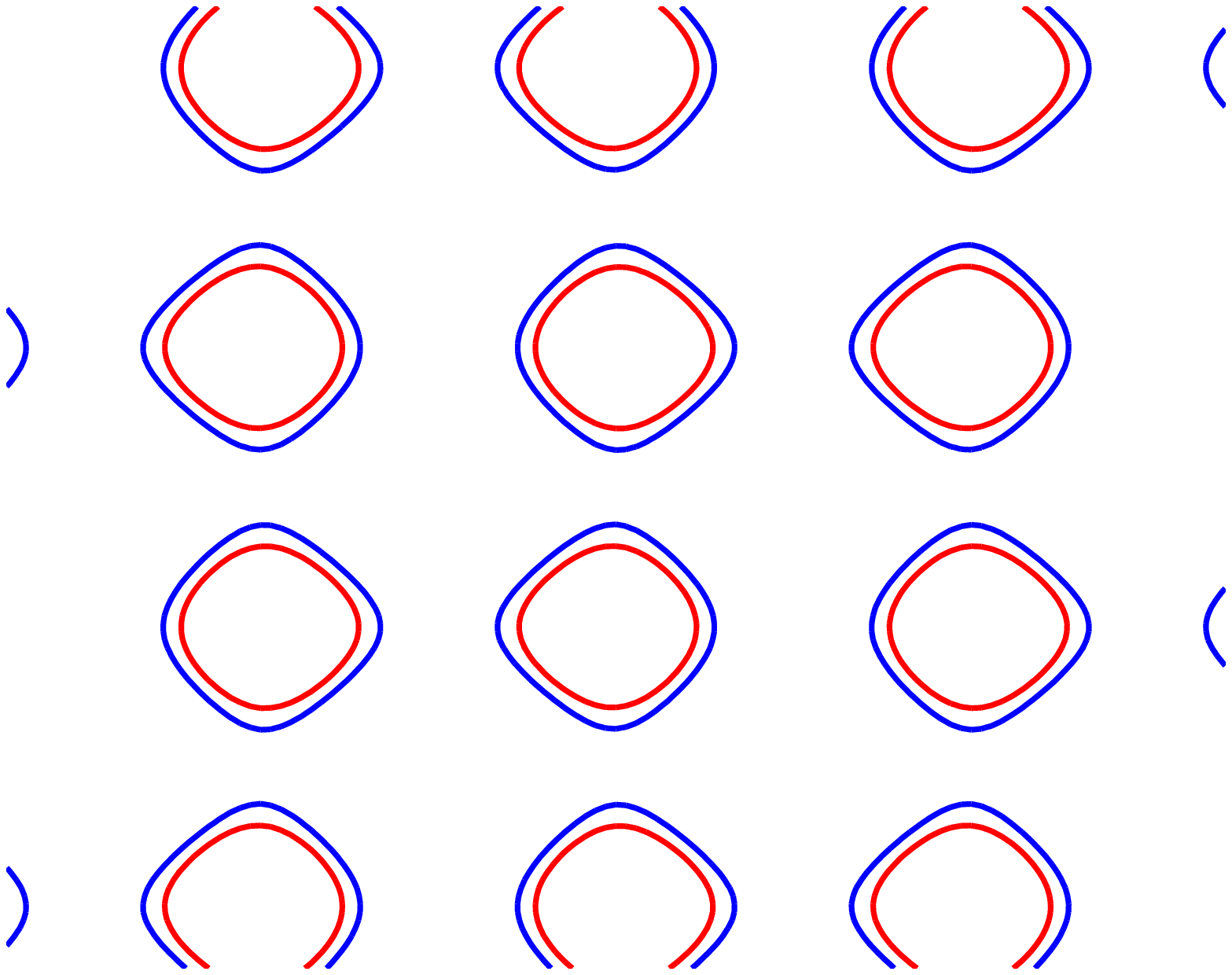}\hfill
\includegraphics[width=7.8cm,clip]{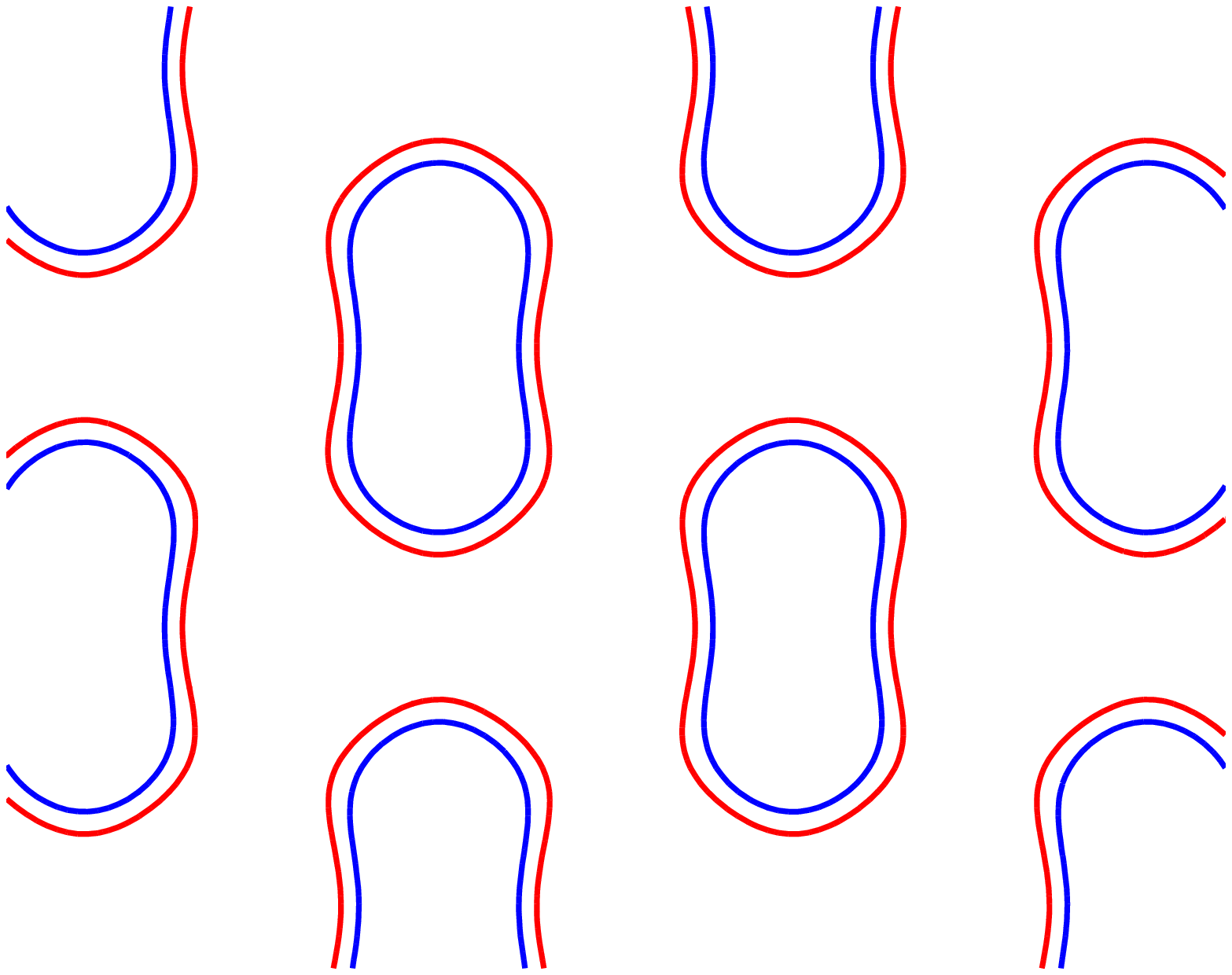}
\caption{\label{fig:pwComp}
Comparison of two source mask parameter combinations leading to vanishing and large process window. (a), (b) Process window; (c),(d) aerial image; (e),(f) developed structure on wafer for intensity levels corresponding to target hole width $\pm 10\%$. Parameters: defocus$=-100\,$nm, $r_1=r_2=0.8$, $b_1=b_2=45^\circ$; (a),(c),(e): $w=192\,$ nm, $h=568\,$nm, $d=104\,$nm; (b),(d),(f): $w=160\,$ nm, $h=488\,$nm, $d=40\,$nm .
}
\end{figure}
As optimization routine we use a combination of global and local minimization algorithms. First we perform a scan with fixed source over the geometrical parameter domain (with $9\cdot 9\cdot 6=486$ forward solutions) and determine mask geometries with large process windows. Figure \ref{fig:pwComp} shows a poor (a,c,e) and the best (b,d,f) result for this run. We nicely observe a narrow waist of the hole pattern in the aerial image in Fig.~\ref{fig:pwComp}(c) for the poor parameters, which is mainly due to the large value of $d$, c.f., Fig.~\ref{fig:target}(b). The target hole height and width can not be printed simultaneously for these parameters. 

For the global geometrical scan, we fixed the source to $r_1=r_2=0.8$ and $b_1=b_2=45^\circ$. The ``good'' geometry candidates now serve as initial guesses for local optimization routines (we use a simplex algorithm, with a maximum number of iterations of 100), which are evoked on the common parameter domain of source and mask. 

The optimal process window is obtained for the following parameter values:
\begin{align}
\begin{split}
r_1&=0.74\\
r_2&=0.89\\
b_1&=43.4^\circ\\
b_2&=44.9^\circ\\
w&=160.0\,\mbox{nm}\\
h&=499.7\,\mbox{nm}\\
d&=39.1\,\mbox{nm}.
\end{split}
\label{eq:optParas}
\end{align}
\begin{figure}[t]
(a)\hspace{8.8cm}(b)\vspace{0.0cm}\\
\includegraphics[width=8.3cm,clip]{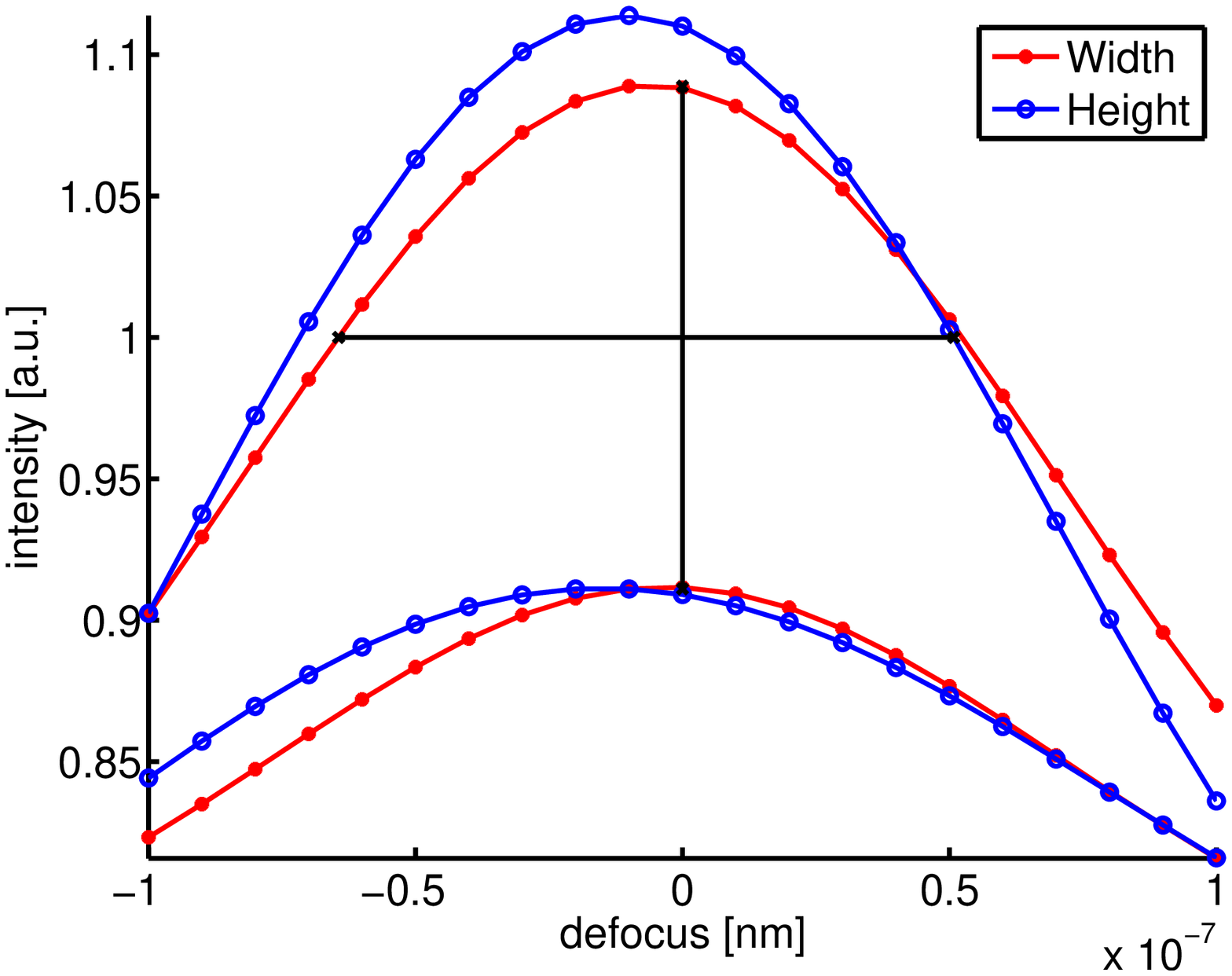}\hfill
\includegraphics[width=8.3cm,clip]{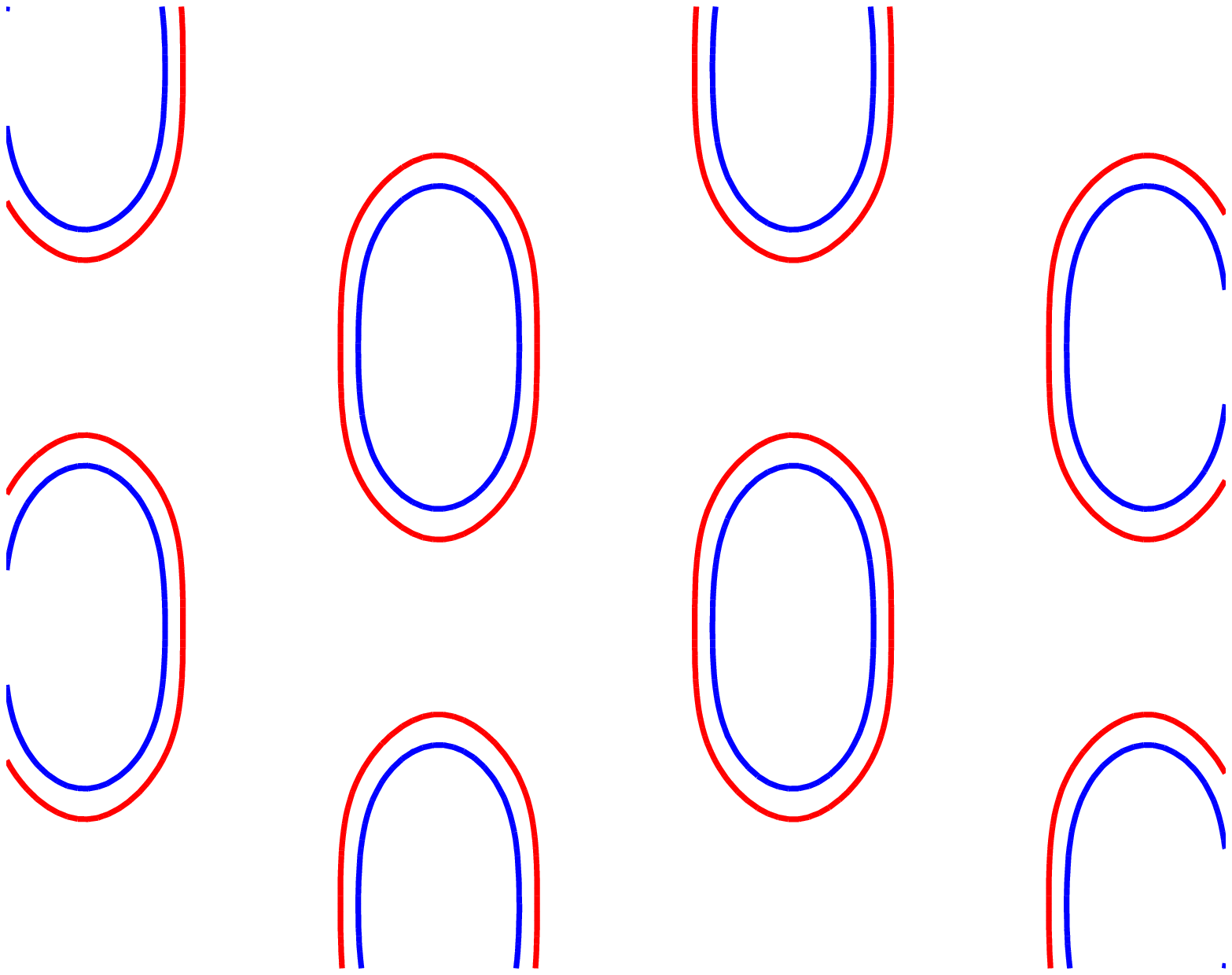}\\
(c)\hspace{8.8cm}(d)\vspace{0.0cm}\\
 \includegraphics[width=8.3cm,clip]{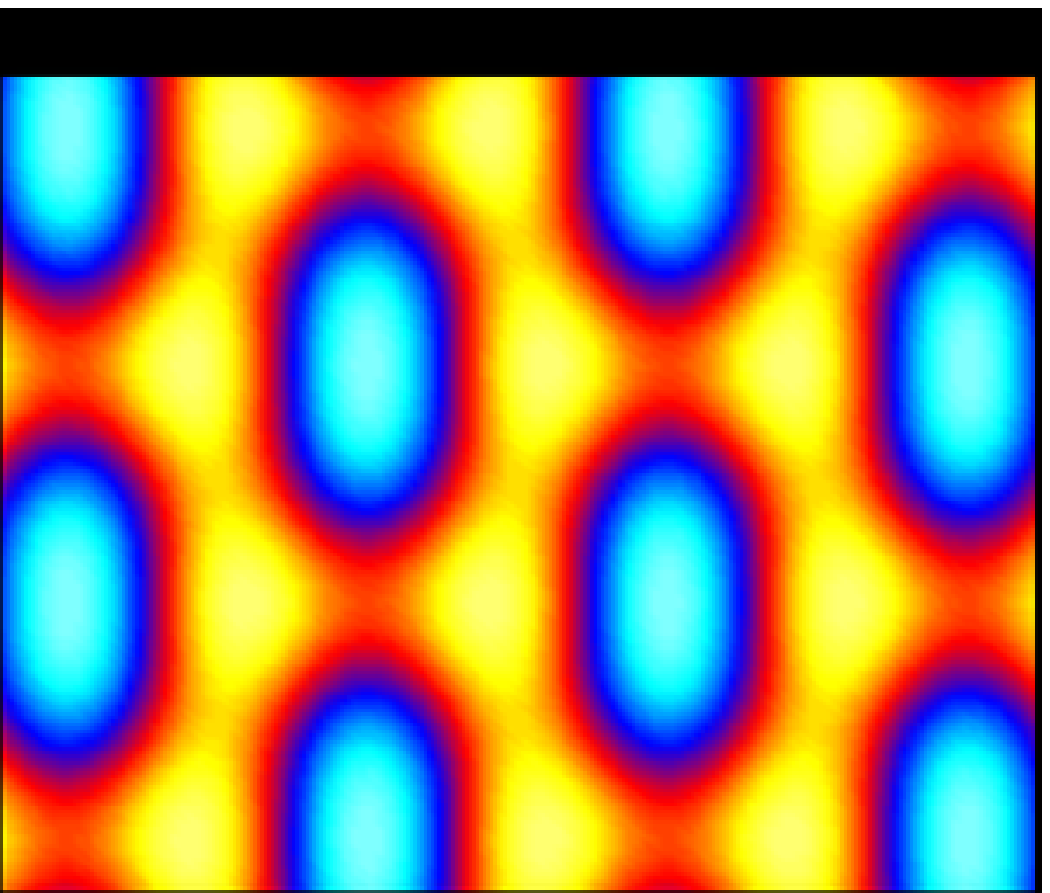}\hfill
 \includegraphics[width=8.3cm,clip]{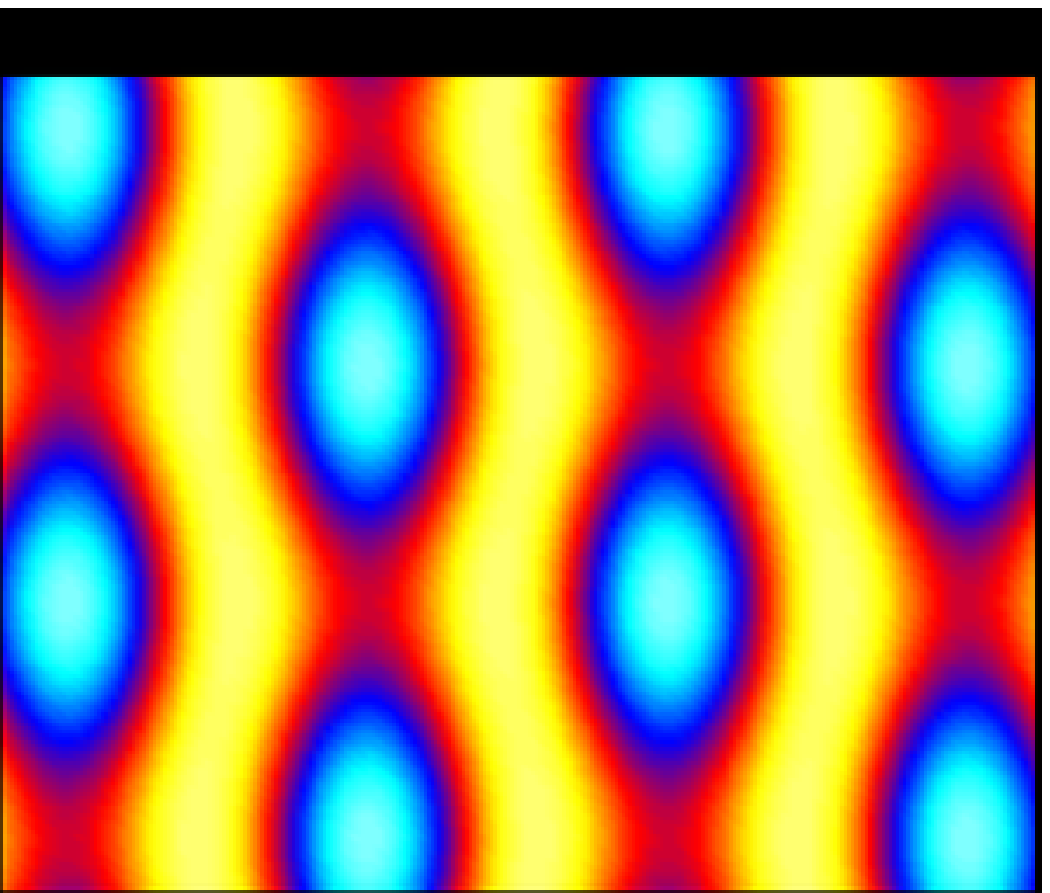}
\caption{\label{fig:pwCompOpt}(a) Optimal process window after source mask optimization; (b) developed structure on wafer at zero defocus for intensity levels corresponding to target hole width $\pm 10\%$. (c),(d) Respective aerial images for zero and $-100\,$nm defocus. Source and mask parameters are given in Equation \eqref{eq:optParas}.
}
\end{figure}
The corresponding process window, aerial images and contours on the wafer are given in Fig.~\ref{fig:pwCompOpt}. Compared to the pure ``geometrical optimization'' in Fig.~\ref{fig:pwComp}(b) we observe that simultaneous optimization of the source further increases the process window. Note that in total we performed roughly 1,000 forward computations for the whole optimization procedure. This is significantly larger than the number of 40 snapshot solutions which were computed in the offline phase for construction of the reduced model. 

The presented source mask optimization was performed on a standard desktop PC (computation time of the order of one hour, including computation of aerial images and structures on the wafer). Without model reduction, the present analysis would have taken over a month. For freeform illumination we expect an even larger number of necessary iterations in an optimization procedure such that offline construction of a reduced basis model pays off even more.

\section{Conclusions}
\label{sec::conclusions}
We presented the reduced basis method as a fast and accurate approach to rigorous solutions of electromagnetic scattering problems in many-query and real-time contexts. In an offline phase the reduced basis method self-adaptively constructs a low dimensional approximation to a given, e.g., geometrically parametrized scattering problem. The reduced system is a scattering problem stated on a reduced basis space and can be solved online very fast in the order of a second. The foundation for construction of the reduced basis approximation is an efficient, rigorous Finite Element solver, e.g., JCMsuite \cite{BUR08}, which was used in the present work. Since the underlying physical model is generically incorporated in the reduced model, exponential convergence rates are observed, which leads to very low dimensional approximations and relatively small construction times of the reduced model. Error estimators furthermore assure the reliability of the reduced basis solutions. 

In a numerical investigation, a reduced model for a geometrically parametrized photomask for cutting hole lithography was constructed and incorporated in a source mask optimization project. The constructed reduced model could be solved about 10,000 times faster than the full problem. A convergence analysis demonstrated that the reduced basis results are in excellent agreement with the exact Finite Element solutions. Furthermore, the good performance of the a posteriori error estimator was demonstrated. Estimated and true errors were highly correlated. Incorporation of the reduced basis technique into the source mask optimization (SMO) process enabled application of combined global and local optimization algorithms for maximization of the process window in a seven-dimensional parameter space. The reduced basis offline phase with only 40 forward solutions thereby paid-off multiply. After construction of the reduced model the optimization with about 1,000 forward solutions could be performed in the order of an hour instead of several weeks, necessary with the full model.


\bibliography{/home/pomplun/myBib}
\bibliographystyle{spiebib}

\end{document}